\documentclass[%
aip,
jcp,
amsmath,amssymb,
preprint,%
]{revtex4-2}

\usepackage{graphicx}
\usepackage{dcolumn}
\usepackage{bm}

\usepackage[utf8]{inputenc}
\usepackage[T1]{fontenc}
\usepackage{mathptmx}
\usepackage{etoolbox}

\usepackage{comment}

\usepackage{enumerate}
\usepackage{algorithm}
\usepackage{algorithmicx}
\usepackage{algpseudocode}
\usepackage{paralist}
\usepackage{threeparttable}
\usepackage{threeparttablex}
\usepackage{booktabs}
\usepackage{makecell}
\usepackage{multirow}
\usepackage{color}
\usepackage[mathscr]{eucal}

\algnewcommand{\Inputs}[1]{
  \State \textbf{Inputs:}
  \Statex \hspace*{\algorithmicindent}\parbox[t]{.8\linewidth}{\raggedright #1}
}
\algnewcommand{\Initialize}[1]{
  \State \textbf{Initialize:}
  \Statex \hspace*{\algorithmicindent}\parbox[t]{.8\linewidth}{\raggedright #1}
}
\algnewcommand{\Outputs}[1]{
  \State \textbf{Outputs:}
  \Statex \hspace*{\algorithmicindent}\parbox[t]{.8\linewidth}{\raggedright #1}
}
\algnewcommand\algorithmicswitch{\textbf{switch}}
\algnewcommand\algorithmiccase{\textbf{case}}
\algdef{SE}[SWITCH]{Switch}{EndSwitch}[1]{\algorithmicswitch\ #1\ \algorithmicdo}{\algorithmicend\ \algorithmicswitch}%
\algdef{SE}[CASE]{Case}{EndCase}[1]{\algorithmiccase\ #1}{\algorithmicend\ \algorithmiccase}%
\algtext*{EndSwitch}%
\algtext*{EndCase}%

\newfloat{Algorithm}{htbp}{alg}
\floatname{Algorithm}{Algorithm}

\makeatletter
\def\@email#1#2{%
 \endgroup
 \patchcmd{\titleblock@produce}
  {\frontmatter@RRAPformat}
  {\frontmatter@RRAPformat{\produce@RRAP{*#1\href{mailto:#2}{#2}}}\frontmatter@RRAPformat}
  {}{}
}%
\makeatother

\UseRawInputEncoding
\begin{document}

\preprint{}

\title[On Committor Functions in Milestoning]{On Committor Functions in Milestoning}
\author{Xiaojun Ji}
\affiliation{Research Center for Mathematics and Interdisciplinary Sciences, Shandong University, Qingdao, Shandong 266237, P. R. China}
\affiliation{Frontiers Science Center for Nonlinear Expectations (Ministry of Education), Shandong University, Qingdao, Shandong 266237, P. R. China}

\author{Ru Wang}
\affiliation{
Qingdao Institute for Theoretical and Computational Sciences, Institute of Frontier and Interdisciplinary Science, Shandong University, Qingdao, Shandong 266237, P. R. China}

\author{Hao Wang*}
\email{wanghaosd@sdu.edu.cn}
\affiliation{
Qingdao Institute for Theoretical and Computational Sciences, Institute of Frontier and Interdisciplinary Science, Shandong University, Qingdao, Shandong 266237, P. R. China}

\author{Wenjian Liu}
\affiliation{
Qingdao Institute for Theoretical and Computational Sciences, Institute of Frontier and Interdisciplinary Science, Shandong University, Qingdao, Shandong 266237, P. R. China}


\begin{abstract}
As an optimal one-dimensional reaction coordinate, the committor function not only describes the probability of a trajectory initiated at a phase space point first reaching the product state before reaching the reactant state, but also preserves the kinetics when utilized to run a reduced dynamics model. However, calculating the committor function in high-dimensional systems poses significant challenges. In this paper, within the framework of Milestoning, exact expressions for committor functions at two levels of coarse graining are given, including committor functions of phase space point to point (CFPP) and milestone to milestone (CFMM). When combined with transition kernels obtained from trajectory analysis, these expressions can be utilized to accurately and efficiently compute the committor functions. Furthermore, based on the calculated committor functions, an adaptive algorithm is developed to gradually refine the transition state region. Finally, two model examples are employed to assess the accuracy of these different formulations of committor functions.
\end{abstract}

\maketitle

\onecolumngrid

\section{Introduction}
In complex systems, transition processes between metastable states, such as protein folding and conformational change, occur on timescales of milliseconds or longer and hence are not easily captured through brute-force molecular dynamics (MD) simulations. 
As such, they are called rare events.
Although rare, their outcomes can significantly impact the functionalities of molecular systems.
Understanding the mechanism underlying these rare events constitutes a central problem in MD simulation studies of complex systems. 

Reaction coordinates provide a simple and valuable picture of transition processes between metastable states.
They are often chosen empirically to capture changes in geometric structure, such as torsional angles, bond distances, and the fraction of native contacts. 
While the selection is usually self-evident in simple chemical reactions, it becomes more intricate for complex biophysical and chemical processes.
Reaction coordinates can also be represented as transition pathways connecting the reactant and product states. 
Numerical methods such as locally updated plane (LUP) method\cite{LUP90}, nudged elastic band (NEB) method\cite{NEB98}, string method\cite{String02,String05}, and transition path sampling (TPS) method\cite{TPSreview02} have been developed to optimize the minimum (free) energy pathway or to sample transition path ensemble. 
Once reaction coordinates are established, the free energy profile can be calculated along the transition process using methods like umbrella sampling\cite{US77}, adaptive biasing force (ABF)\cite{ABF01,ABF15}, and metadynamics\cite{Metadynamics02}. 
Furthermore, kinetic properties such as rate constants can be computed using methods such as transition interface sampling (TIS)\cite{TIS03}, weighted ensemble (WE)\cite{WE96,WE10}, forward flux sampling (FFS)\cite{FFS09}, and Milestoning\cite{Milestoning04}.  

A single reaction coordinate is appealing due to its simplicity. 
The committor function is considered an optimal one-dimensional reaction coordinate\cite{OptRC16,CommOptFolding18} in the sense that it preserves the kinetics, such as reaction flux\cite{OptRC13,DiffComm13} and mean first passage time (MFPT)\cite{ExCG14}, when employed to run a reduced dynamics model. 
In particular, iso-committor surfaces are optimal milestones (hypersurfaces) for exact MFPT calculations within the framework of Milestoning\cite{OptMilestone08}.
The committor function is defined in configuration or phase space, describing the probability of a trajectory initiated at a (phase space) configuration first reaching the product state before reaching the reactant state. 
The transition state region is indicated by committor values around $1/2$. 
Based on the committor function, equilibrium probability distribution, probability current, and transition pathways can be analyzed, as illustrated in the transition path theory (TPT)\cite{TPT06,TPT09,TPT10}.

In overdamped Langevian dynamics, the committor function satisfies the backward Kolmogorov equation\cite{StoCal96}. 
However, directly solving this equation for the committor function can be quite expensive. 
Conventional numerical techniques such as the finite elements method are limited to solving the equation in two dimensions or three dimensions. 
Recently, neural networks are utilized to solve the equation for up to ten dimensions in model systems\cite{CommNN18}, where solving the high-dimensional partial differential equation has been transformed to a variational problem. 
Following the same variational formulation, an alternative practical method is developed for complex systems, where approximations that the committor function can be expressed as a function of a few collective variables have to be made\cite{CommNN19}.

The committor function can also be calculated by trajectory analysis. 
In the case when a single equilibrium long-enough trajectory simulation is available, e.g., from the special-purpose molecular simulation machine\cite{Anton,Shaw11}, the committor function can be determined by an adaptive nonparametric optimization approach via minimizing the total squared displacement\cite{Krivov15}. 
See Ref. \cite{CommOptFolding18} for a successful application in protein folding dynamics. 
However, a single equilibrium long-enough trajectory is not always affordable, especially for complex systems. 
In contrast, a more efficient approach is to conduct simulations using short trajectory ensemble, from which the transition kernel among discretized states can be obtained. 
The committor function values on these discretized states are connected though the transition kernel.
Within the framework of Milestoning, the committor function can be computed based on an assumption of constant committor values on milestones\cite{CommMilestoning17}. 
This assumption is justified when milestones are deployed along iso-committor surfaces, or when the size of the milestones is sufficiently small, thus rendering the fluctuations in committor function values on milestones negligible.
However, this assumption does not hold true in general cases when milestones are arbitrarily deployed.

In this paper, exact expressions for committor functions at two levels of coarse graining are derived within the framework of Milestoning. 
These expressions include committor functions of phase space point to point (CFPP) and milestone to milestone (CFMM) and can be applied to arbitrarily deployed milestones. 
The CFPP provides a detailed characterization of the transition process but is computationally expensive, as it requires a comprehensive transition kernel.
On the other hand, the CFMM, which is the average committor value, provides a coarser picture and can be efficiently calculated with the help of the transition probability matrix within the low-dimensional milestone state space.

The remainder of this paper is organized as follows. 
First in Sec. \ref{Milestoning} the Milestoning theory is briefly reviewed. 
Next, in Sec. \ref{Comm Func}, exact expressions for committor functions at two levels of coarse graining are derived. 
In Sec. \ref{FHPD Comm}, simplified expressions for the CFMM are given. 
Finally, in Sec. \ref{Results}, two model examples are employed to assess the accuracy of these different formulations of committor functions.

\section{Method}\label{Methods}
\subsection{Milestoning Backdrop}\label{Milestoning}
A system of $N$ particles at time $t$ is fully characterized by a phase space point, $x(t)$.
The motion in the full phase space is assumed to be a Markov process. This assumption remains valid for prevalent MD simulations such as Hamiltonian dynamics and Langevin dynamics.

\begin{figure}[h]
\centering
\begin{tabular}{cc}
\includegraphics[height=7cm]{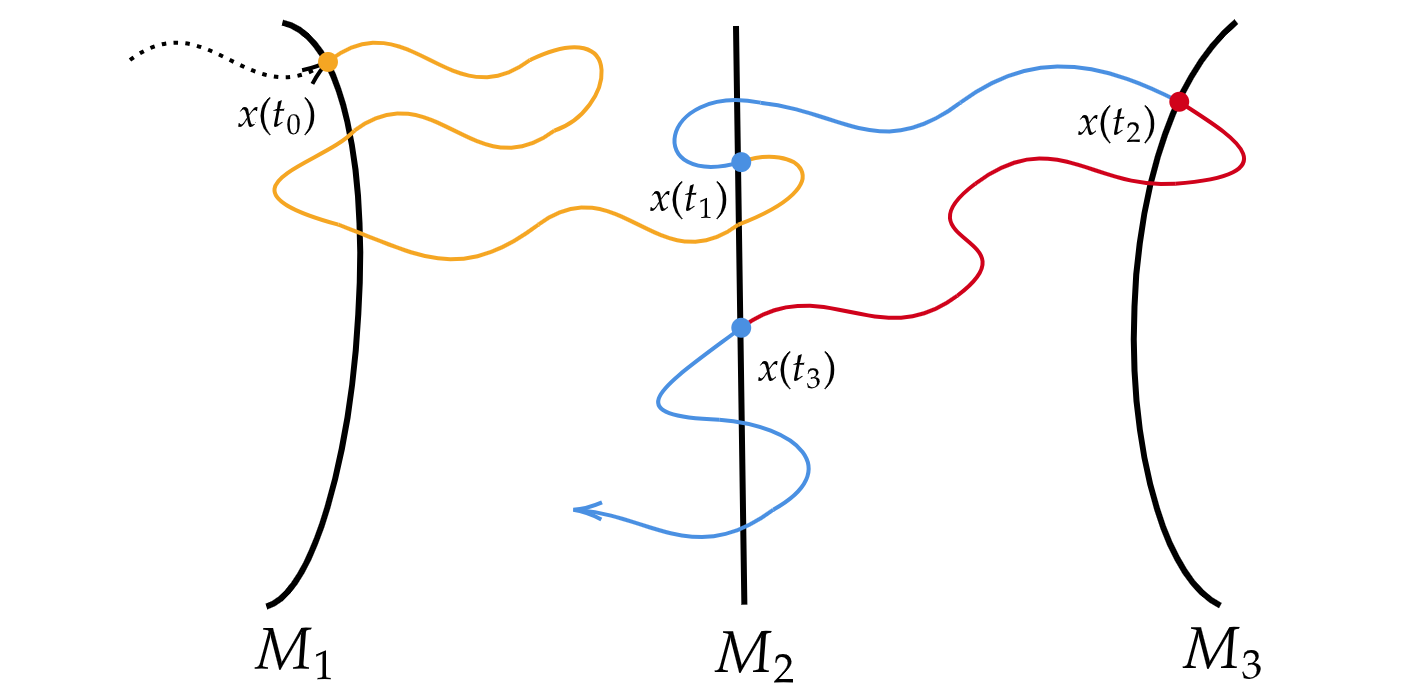}\\
(a)\\
\includegraphics[height=7cm]{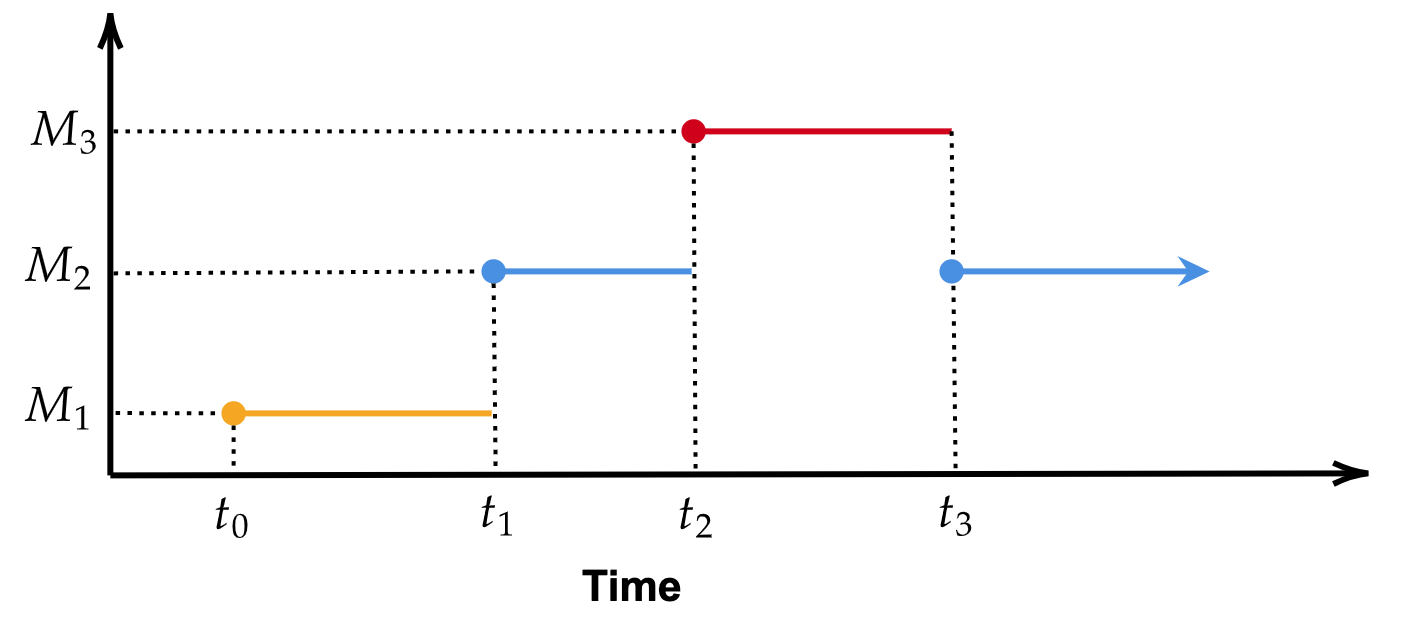}\\
(b)
\end{tabular}
\caption{(a) A segment from an equilibrium long trajectory passing through three consecutive milestones. The current state of the trajectory, coded in different colors, is determined by the last milestone it crossed. First hitting points are depicted as dots. (b) The stochastic transition of the long trajectory mapped into the discretized milestone state space.}\label{fig_milestoning}
\end{figure}

In Milestoning, the full phase space is partitioned into small compartments using techniques like Voronoi tessellation\cite{MarkovMile}.
Interfaces between compartments are called milestones, denoted as $\{M_1,\cdots,M_n\}$.
Fluxes through milestones are integrated to calculate thermodynamic and kinetic properties, such as the free energy profile and MFPT\cite{Milestoning04,ExM15,LPTM23}.
At the core of Milestoning lies the fundamental event of transiting from a first hitting point $x_\alpha$ on milestone $\alpha$ to another first hitting point $x_\beta$ on milestone $\beta$, $\alpha,\beta\in\{M_1,\cdots,M_n\}$ (cf. Fig. \ref{fig_milestoning}). 
A first hitting point on a milestone is a phase space point that a trajectory initially reaches upon after crossing a different milestone.
The subsequent discussions concerning crossing events on milestones pertain specifically to the concept of crossing as a first hitting point.
The current state of the trajectory is determined by the last milestone it crossed, and recrossing the same milestone does not alter the current trajectory state. 
It is worth noting that the stochastic transition in the low-dimensional discretized milestone state space is non-Markovian.

Let us first introduce three important functions that are utilized extensively in Milestoning.
The first is the probability density, $q_\alpha(x_\alpha,t)$, that milestone $\alpha$ is being crossed at the point $x_\alpha$ and time $t$. 
It is noteworthy that $q_\alpha(x_\alpha,t)$ also represents the flux at the point $x_\alpha$ and time $t$.
The second is the probability density, $p_\alpha(x_\alpha,t)$, that the last crossing event is situated at $x_\alpha$ on milestone $\alpha$ and time $t'$, $t'<t$.
The third is the transition kernel $K_{\beta\alpha}(x_\beta,t',x_\alpha,t)$, which is defined as the conditional probability density of crossing milestone $\alpha$ at $x_\alpha$ and time $t$, given that milestone $\beta$ was crossed at $x_\beta$ and time $t'$, where $t'<t$. No other milestones are crossed in the time interval $[t',t]$.

The flux functions at different milestones are connected via the transition kernel\cite{ExM15},
\begin{equation}
q_\alpha(x_\alpha,t)=p_\alpha(x_\alpha,t=0)\delta(t)+\sum_{\beta\neq\alpha}\int_0^t dt'\int_\beta dx_\beta\ q_\beta(x_\beta,t')K_{\beta\alpha}(x_\beta,t',x_\alpha,t),
\label{flux cons}
\end{equation}
where $\delta(t)$ is the Dirac delta function and $p_\alpha(x_\alpha,t=0)$ is the initial distribution on milestone $\alpha$. 
The restriction imposed on summation, namely $\beta\neq\alpha$, can be omitted. This is due to the fact that $K_{\beta\beta}(x_\beta^{(1)},t',x_\beta^{(2)},t)$, denoting a recrossing event, is inherently defined as zero.

Eq. \eqref{flux cons} plays a central role in Milestoning, expressing the fundamental conservation of flux due to our presumption of the absence of sources or sinks in the phase space. 
The left-hand side of the equation is the probability (density) of milestone $\alpha$ being crossed at $x_\alpha$ and time $t$. 
At the initial time $t=0$, this probability aligns with the initial distribution on milestone $\alpha$, $p_\alpha(x_\alpha,t=0)$. 
As time progresses $t>0$, it is equal to the probability of crossing a neighboring milestone $\beta$ at $x_\beta$ and an earlier time $t'<t$, multiplied by the conditional transition probability from $x_\beta$ to $x_\alpha$ precisely at time $t$. 
Here, the Markov property in the full phase space is employed. 
Finally, different transition pathways are summed up.  

When there is no time-dependent external perturbation, as is the case under consideration in this paper, the transition kernel is time-homogeneous. 
This implies that $K_{\beta\alpha}(x_\beta,t',x_\alpha,t)=K_{\beta\alpha}(x_\beta,x_\alpha,t-t')$. Consequently, the second term on the right-hand side of Eq. \eqref{flux cons} takes on a convolutional form.

After a sufficiently long period of time, the system will attain a stationary state, where the flux through each milestone no longer depends on time, $q_{\mathrm{stat},\alpha}(x_\alpha)=\lim_{t\rightarrow \infty}q_\alpha(x_\alpha,t)$. 
The stationary flux is related to the Laplace transform of $q_\alpha(x_\alpha,t)$ in the limit $u\rightarrow 0$, represented as $q_{\mathrm{stat},\alpha}(x_\alpha)=\lim_{u\rightarrow 0}u\tilde{q}_\alpha(x_\alpha,u)$. This relationship can be verified as follows,
\begin{align}
\lim_{u\rightarrow 0}u\tilde{q}_\alpha(x_\alpha,u) &= \lim_{u\rightarrow 0}u\int_0^\infty dt\ e^{-ut}q_\alpha(x_\alpha,t)\nonumber\\
&=\lim_{u\rightarrow 0}u[\int_0^\tau dt\ e^{-ut}q_\alpha(x_\alpha,t)+q_{\mathrm{stat},\alpha}(x_\alpha)\int_\tau^\infty dt\ e^{-ut}]\nonumber\\
&=0+q_{\mathrm{stat},\alpha}(x_\alpha)\lim_{u\rightarrow 0}u\cdot\frac{1}{u}e^{-u\tau}\nonumber\\
&=q_{\mathrm{stat},\alpha}(x_\alpha),
\end{align}
where $\tau$ is a time point of sufficient magnitude after which the flux remains stationary.

This observation inspires us to take the Laplace transform on both sides of Eq. \eqref{flux cons},
\begin{equation}
\tilde{q}_\alpha(x_\alpha,u)=p_\alpha(x_\alpha,t=0)+\sum_{\beta}\int_\beta dx_\beta\ \tilde{q}_\beta(x_\beta,u)\tilde{K}_{\beta\alpha}(x_\beta,x_\alpha,u).
\label{flux cons laplace}
\end{equation}
Multiplying both sides of Eq. \eqref{flux cons laplace} by $u$ and taking the limit $u\rightarrow 0$, we obtain
\begin{equation}
q_{\mathrm{stat},\alpha}(x_\alpha) = \sum_\beta\int_\beta dx_\beta\ q_{\mathrm{stat},\beta}(x_\beta)K_{\beta\alpha}(x_\beta,x_\alpha),
\label{q stat}
\end{equation}
where we denote $K_{\beta\alpha}(x_\beta,x_\alpha)=\lim_{u\rightarrow 0}\tilde{K}_{\beta\alpha}(x_\beta,x_\alpha,u)=\int_0^\infty dt\ K_{\beta\alpha}(x_\beta,x_\alpha,t)$. 
The term $K_{\beta\alpha}(x_\beta,x_\alpha)$ is the transition probability density from $x_\beta$ to $x_\alpha$ regardless of the transition time.

\subsection{Committor Functions}\label{Comm Func}
To define committor functions, two end states need to be first designated. 
These two end states, $\theta$ and $\phi$, are designated in the milestone state space and are regarded as the reactant and product states, respectively.
Let us begin by calculating the CFPP, $C(x_\phi; x_\zeta)$, representing the probability of a trajectory initiated at $x_\zeta$ on an intermediate milestone $\zeta$ first reaching $x_\phi$ on the product milestone $\phi$ before reaching the reactant milestone $\theta$. 
The initial distribution is prepared on milestone $\zeta$,
\begin{equation}
p_\alpha(x_\alpha,t=0)=\begin{cases}
\rho_\zeta(x_\zeta) & \mathrm{if}\ \alpha = \zeta, \\
 0 & \mathrm{if}\ \alpha\neq \zeta, \\
\end{cases}
\label{initial distribution}
\end{equation}
where $\rho_\zeta(x_\zeta)$ is the prescribed normalized initial distribution on milestone $\zeta$.
Given the initial distribution on milestone $\zeta$, the average committor value of milestone $\zeta$ to $x_\phi$ is calculated as
\begin{equation}
C_\zeta(x_\phi) = \int_\zeta dx_\zeta\ \rho_\zeta(x_\zeta)C(x_\phi; x_\zeta).
\label{comm to point}
\end{equation}
When the initial distribution on milestone $\zeta$ takes the form of a delta function, $\rho_\zeta(x_\zeta)=\delta(x_\zeta-x_\zeta')$, Eq. \eqref{comm to point} recovers the CFPP $C(x_\phi; x_\zeta')$.

To facilitate the calculation, absorbing boundary conditions are imposed at two end milestones $\theta$ and $\phi$, 
\begin{equation}
K^{(A)}_{\beta\alpha}(x_\beta,x_\alpha)=\begin{cases}
K_{\beta\alpha}(x_\beta,x_\alpha) &\mathrm{if}\ \beta\neq\theta\ \mathrm{and}\ \phi,\\
0 & \mathrm{if}\ \beta=\theta\ \mathrm{or}\ \phi.
\end{cases}
\end{equation}
This approach permits the convenient calculation of $C_\zeta(x_\phi)$ via the integration of flux at milestone $\phi$, i.e.,
\begin{equation}
C_\zeta(x_\phi) = \int_0^\infty dt\ q_\phi(x_\phi,t).
\label{C by q}
\end{equation}

Note that the right-hand side of Eq. \eqref{C by q} can also be regarded as the Laplace transform of $q_\phi(x_\phi,t)$ in the limit $u\rightarrow 0$,
\begin{equation}
\int_0^\infty dt\ q_\phi(x_\phi,t)=\lim_{u\rightarrow 0}\int_0^\infty dt\ e^{-ut}q_\phi(x_\phi,t)=\lim_{u\rightarrow 0}\tilde{q}_\phi(x_\phi,u).
\label{q laplace}
\end{equation}

Taking directly the limit $u\rightarrow 0$ on both sides of Eq. \eqref{flux cons laplace}, we obtain
\begin{equation}
C'_\zeta(x_\alpha) = p_\alpha(x_\alpha,t=0) + \sum_{\beta}\int_\beta dx_\beta\ C'_\zeta(x_\beta)K_{\beta\alpha}^{(A)}(x_\beta,x_\alpha),
\label{pseudo comm eq}
\end{equation}
where we denote $C'_\zeta(x_\alpha)=\lim_{u\rightarrow 0}\tilde{q}_\alpha(x_\alpha,u)$ and impose absorbing boundary conditions at two end milestones $\theta$ and $\phi$ in the transition kernel. 
Note that $C'_\zeta(x_\alpha)$ does not represent the average committor value of milestone $\zeta$ to $x_\alpha$ for an arbitrary milestone $\alpha$. 
It bears the meaning of the average committor value only when $\alpha=\phi$. 

Eq. \eqref{pseudo comm eq} can be reformulated as a operator equation of simpler form,
\begin{equation}
\mathbf{I}_dC_\zeta'=p(t=0)+\mathcal{K}^{(A)}C_\zeta',
\label{C' eq}
\end{equation}
where $\mathcal{K}^{(A)}$ is a linear transition operator corresponding to the transition kernel under absorbing boundary conditions,
\begin{equation}
(\mathcal{K}^{(A)}C_\zeta')(x_\alpha)\equiv\sum_{\beta}\int_\beta dx_\beta\ C'_\zeta(x_\beta)K_{\beta\alpha}^{(A)}(x_\beta,x_\alpha),
\end{equation}
$\mathbf{I}_d$ is an identity operator acting on $C_\zeta'$, 
\begin{equation}
(\mathbf{I}_dC_\zeta')(x_\alpha) \equiv C_\zeta'(x_\alpha),
\end{equation}
and the initial distribution is denoted by
\begin{equation}
p(t=0)(x_\alpha)=p_\alpha(x_\alpha,t=0).
\end{equation}
Rearranging Eq. \eqref{C' eq} for $C_\zeta'$, we obtain
\begin{equation}
C_\zeta'=(\mathbf{I}_d-\mathcal{K}^{(A)})^{-1}p(t=0).
\label{Czeta'}
\end{equation}

Based on the expression obtained for $C_\zeta'$, the average committor value $C_\zeta(x_\phi)$ of milestone $\zeta$ to $x_\phi$ (which covers the CFPP $C(x_\phi;x_\zeta)$ as a special case) can be computed by
\begin{equation}
C_\zeta(x_\phi) = \mathcal{I}_{x_\phi}C_\zeta'=\mathcal{I}_{x_\phi}(\mathbf{I}_d-\mathcal{K}^{(A)})^{-1}p(t=0),
\label{comm from C'}
\end{equation}
where $\mathcal{I}_{x_\phi}$ is a linear operator that evaluates a function on $x_\phi$ for any given $x_\phi\in\phi$,
\begin{equation}
\mathcal{I}_{x_\phi}C_\zeta'\equiv\sum_\beta\int_\beta dx_\beta\ C_\zeta'(x_\beta)\delta(x_\beta-x_\phi)=C_\zeta'(x_\phi).
\end{equation}
Additionally, the CFMM $C_\zeta$, which represents the average committor value of milestone $\zeta$ to $\phi$, can be evaluated as well,
\begin{equation}
C_\zeta =\mathcal{I}_\phi C_\zeta' = \mathcal{I}_\phi(\mathbf{I}_d-\mathcal{K}^{(A)})^{-1}p(t=0),
\label{comm to milestone matrix}
\end{equation}
where $\mathcal{I}_\phi$ is a linear operator that evaluates the integral of a function over milestone $\phi$,
\begin{equation}
\mathcal{I}_\phi C_\zeta'\equiv\sum_\beta\int_\beta dx_\beta\ C_\zeta'(x_\beta)\mathbb{I}_{x_\beta\in\phi}=\int_\phi dx_\phi\ C_\zeta'(x_\phi),
\end{equation}
with $\mathbb{I}_{x_\beta\in\phi}$ being an indicator function which takes value 1 when $x_\beta$ is on milestone $\phi$ and 0 otherwise.

A simpler expression of $C_\zeta'$ can be obtained by transforming the transition kernel into cyclic boundary conditions at two end milestones $\theta$ and $\phi$,
\begin{equation}
K_{\beta\alpha}^{(C)}(x_\beta,x_\alpha)=\begin{cases}
K_{\beta\alpha}(x_\beta,x_\alpha) & \mathrm{if}\ \beta\neq\theta\ \mathrm{and}\ \phi,\\
p(t=0)(x_\alpha) & \mathrm{if}\ \beta=\theta\ \mathrm{or}\ \phi.\\
\end{cases}
\end{equation}
The transition kernels under absorbing boundary conditions and cyclic boundary conditions are connected by
\begin{equation}
K^{(A)}_{\beta\alpha}(x_\beta,x_\alpha) = K^{(C)}_{\beta\alpha}(x_\beta,x_\alpha) - p(t=0)(x_\alpha)(\mathbb{I}_{\beta=\theta}+\mathbb{I}_{\beta=\phi}).
\label{AC point}
\end{equation}

To proceed, we introduce a linear transition operator $\mathcal{K}^{(C)}$ corresponding to the transition kernel under cyclic boundary conditions,
\begin{equation}
(\mathcal{K}^{(C)}q)(x_\alpha)\equiv\sum_\beta\int_\beta dx_\beta\ q_\beta(x_\beta)K_{\beta\alpha}^{(C)}(x_\beta,x_\alpha),
\end{equation}
and assume $q_{\mathrm{stat}}$ is the stationary solution of $\mathcal{K}^{(C)}$ (cf. Eq. \eqref{q stat}), 
\begin{equation}
\mathcal{K}^{(C)}q_{\mathrm{stat}}=q_{\mathrm{stat}}.
\label{q stat matrix}
\end{equation}
Noting that
\begin{equation}
\mathcal{K}^{(A)}=\mathcal{K}^{(C)}-p(t=0)(\mathcal{I}_\theta+\mathcal{I}_\phi),
\end{equation}
we obtain the following equality
\begin{align}
(\mathbf{I}_d-\mathcal{K}^{(A)})q_{\mathrm{stat}} &= q_{\mathrm{stat}}-\mathcal{K}^{(A)}q_{\mathrm{stat}}\nonumber\\
&=q_{\mathrm{stat}}-\mathcal{K}^{(C)}q_{\mathrm{stat}}+p(t=0)(\mathcal{I}_\theta+\mathcal{I}_\phi)q_{\mathrm{stat}}\nonumber\\
&=p(t=0)(\mathcal{I}_\theta+\mathcal{I}_\phi)q_{\mathrm{stat}}.
\label{operator identity}
\end{align}
Multiplying both sides of Eq. \eqref{Czeta'} by a scalar $(\mathcal{I}_\theta+\mathcal{I}_\phi)q_{\mathrm{stat}}$, we obtain
\begin{align}
C_\zeta'(\mathcal{I}_\theta+\mathcal{I}_\phi)q_{\mathrm{stat}} &= (\mathbf{I}_d-\mathcal{K}^{(A)})^{-1}p(t=0)(\mathcal{I}_\theta+\mathcal{I}_\phi)q_{\mathrm{stat}}\nonumber\\
&= (\mathbf{I}_d-\mathcal{K}^{(A)})^{-1}(\mathbf{I}_d-\mathcal{K}^{(A)})q_{\mathrm{stat}}\nonumber\\
&= q_{\mathrm{stat}},
\label{comm cyc}
\end{align}
where the second equality comes from Eq. \eqref{operator identity}.
Rearrange Eq. \eqref{comm cyc} to obtain the final expression for $C_\zeta'$
\begin{equation}
C_\zeta' = \frac{q_{\mathrm{stat}}}{(\mathcal{I}_\theta+\mathcal{I}_\phi)q_{\mathrm{stat}}}.
\label{comm cyc final Czeta'}
\end{equation}
As a result, the average committor value $C_\zeta(x_\phi)$ of milestone $\zeta$ to $x_\phi$ is given by
\begin{align}
C_\zeta(x_\phi) &= \frac{\mathcal{I}_{x_\phi}q_{\mathrm{stat}}}{(\mathcal{I}_\theta+\mathcal{I}_\phi)q_{\mathrm{stat}}} \nonumber\\
&=\frac{q_{\mathrm{stat},\phi}(x_\phi)}{\int_\theta dx_\theta\ q_{\mathrm{stat},\theta}(x_\theta)+\int_\phi dx_\phi\ q_{\mathrm{stat},\phi}(x_\phi)},
\label{Czeta xphi cyc}
\end{align}
and the CFMM $C_\zeta$ is given by
\begin{align}
C_\zeta &= \frac{\mathcal{I}_\phi q_{\mathrm{stat}}}{(\mathcal{I}_\theta+\mathcal{I}_\phi)q_{\mathrm{stat}}} \nonumber\\
&=\frac{\int_\phi dx_\phi\ q_{\mathrm{stat},\phi}(x_\phi)}{\int_\theta dx_\theta\ q_{\mathrm{stat},\theta}(x_\theta)+\int_\phi dx_\phi\ q_{\mathrm{stat},\phi}(x_\phi)}.
\label{comm cyc final}
\end{align}

In practice, one possible way to solve the operator equations \eqref{comm to milestone matrix} or \eqref{q stat matrix} is to perform mesh grid discretization on each milestone as illustrated in Sec. \ref{Results}. 
However, this procedure is generally expensive, as it requires an inversion or solving the eigen-equation of a comprehensive transition probability matrix, whose size is usually very large for complex systems. 
Nonetheless, Eq. \eqref{comm cyc final} inspires us a more computationally efficient approach. See Sec. \ref{FHPD Comm} below.

\subsection{Simplified Expressions for the CFMM}\label{FHPD Comm}
At stationary state, the flux through an arbitrary milestone $\alpha$ can be decomposed into a product of the total flux $w_\alpha$ and a normalized first hitting point distribution (FHPD) $f_\alpha(x_\alpha)$,
\begin{equation}
q^\zeta_{\mathrm{stat},\alpha}(x_\alpha)=w^\zeta_\alpha f^\zeta_\alpha(x_\alpha), 
\label{q stat decomp}
\end{equation}
\begin{equation}
\int_\alpha dx_\alpha\ f^\zeta_\alpha(x_\alpha)=1,
\end{equation}
for any $\alpha\in\{M_1,\cdots,M_n\}$. The superscript $\zeta$ indicates the implicit dependence of the stationary flux on the initial distribution prepared on milestone $\zeta$.

By substituting Eq. \eqref{q stat decomp} into Eq. \eqref{q stat} and integrating over $x_\alpha$ on both sides, we obtain
\begin{equation}
w^\zeta_\alpha = \sum_\beta w^\zeta_\beta\bar{K}^\zeta_{\beta\alpha},
\label{w stat eq}
\end{equation}
where the transition probability matrix $\bar{\mathbf{K}}^\zeta$ is defined as
\begin{equation}
\bar{K}^\zeta_{\beta\alpha} = \int_\beta dx_\beta\int_\alpha dx_\alpha\ f^\zeta_\beta(x_\beta)K_{\beta\alpha}(x_\beta,x_\alpha).
\label{averaged transit prob}
\end{equation}
Note that the transition kernel $K_{\beta\alpha}(x_\beta,x_\alpha)$ does not depend on the initial distribution.

In practice, the transition probability $\bar{K}^\zeta_{\beta\alpha}$ can be calculated by initiating $n_\beta$ trajectories on milestone $\beta$ in accordance with the FHPD, $f^\zeta_\beta(x_\beta)$. 
Subsequently, the number of these trajectories that first reach milestone $\alpha$ is counted and denoted as $n_{\beta\alpha}$, leading to $\bar{K}^\zeta_{\beta\alpha}=n_{\beta\alpha}/n_\beta$. 
Various algorithms are available for approximating the FHPD using parallel short trajectory simulations, offering different levels of accuracy and efficiency\cite{CM04,ExM15,LPTM23}.

Rewrite Eq. \eqref{w stat eq} into a matrix form, incorporating cyclic boundary conditions at two end milestones $\theta$ and $\phi$,
\begin{equation}
(\mathbf{w}^\zeta)^T=(\mathbf{w}^\zeta)^T\bar{K}^{(C),\zeta}.
\label{w stat matrix}
\end{equation}
Eq. \eqref{comm cyc final} can then be expressed in terms of $\mathbf{w}^\zeta$,
\begin{equation}
C_\zeta = \frac{w^\zeta_\phi}{w^\zeta_\theta+w^\zeta_\phi}.
\label{comm cyc w}
\end{equation}
Eq. \eqref{w stat matrix} combined with Eq. \eqref{comm cyc w} constitutes an efficient way of calculating the CFMM $C_\zeta$.

Analogous to Eq. \eqref{AC point}, a transition probability matrix accounting for absorbing boundary conditions at two end milestones $\theta$ and $\phi$ can be defined,
\begin{equation}
\bar{\mathbf{K}}^{(A),\zeta}=\bar{\mathbf{K}}^{(C),\zeta}-(\mathbf{e}_\theta+\mathbf{e}_\phi)(\mathbf{p}(t=0))^T,
\label{matrix AC connect}
\end{equation}
where $\mathbf{e}_{\theta}$ (or $\mathbf{e}_\phi$) is a unit vector of length $n$, $\mathbf{e}_{\cdot}=(0,\cdots,0,1,0,\cdots,0)^T$ with the nonzero element corresponding to milestone $\theta$ (or $\phi$), and $\mathbf{p}(t=0)=\mathbf{e}_\zeta$. 
That is, the transition probability matrices $\bar{\mathbf{K}}^{(A),\zeta}$ and $\bar{\mathbf{K}}^{(C),\zeta}$ only differ in two rows corresponding to two end milestones $\theta$ and $\phi$. 

To obtain a simplified expression for $C_\zeta$ using the transition probability matrix under absorbing boundary conditions, we observe the following identity
\begin{align}
(\mathbf{w}^\zeta)^T(\mathbf{I}-\bar{\mathbf{K}}^{(A),\zeta}) &= (\mathbf{w}^\zeta)^T-(\mathbf{w}^\zeta)^T\bar{\mathbf{K}}^{(A),\zeta}\nonumber\\
&=(\mathbf{w}^\zeta)^T-(\mathbf{w}^\zeta)^T(\bar{\mathbf{K}}^{(C),\zeta}-(\mathbf{e}_\theta+\mathbf{e}_\phi)(\mathbf{p}(t=0))^T)\nonumber\\
&=(\mathbf{w}^\zeta)^T-(\mathbf{w}^\zeta)^T\bar{\mathbf{K}}^{(C),\zeta}+(w^\zeta_\theta+w^\zeta_\phi)(\mathbf{p}(t=0))^T\nonumber\\
&=(w^\zeta_\theta+w^\zeta_\phi)(\mathbf{p}(t=0))^T,
\label{matrix absorb identity}
\end{align}
where going from the first line to the second line we use Eq. \eqref{matrix AC connect} and going from the third line to the last line we use Eq. \eqref{w stat matrix}.
Rearranging Eq. \eqref{matrix absorb identity} into 
\begin{equation}
(\mathbf{w}^\zeta)^T=(w^\zeta_\theta+w^\zeta_\phi)(\mathbf{p}(t=0))^T(\mathbf{I}-\bar{\mathbf{K}}^{(A),\zeta})^{-1},
\end{equation}
and multiplying a column vector $\mathbf{e}_\phi$ to the right on both sides, we arrive at
\begin{equation}
w^\zeta_\phi = (\mathbf{w}^\zeta)^T\mathbf{e}_\phi=(w^\zeta_\theta+w^\zeta_\phi)(\mathbf{p}(t=0))^T(\mathbf{I}-\bar{\mathbf{K}}^{(A),\zeta})^{-1}\mathbf{e}_\phi.
\end{equation}
Comparing with Eq. \eqref{comm cyc w}, we obtain an alternative expression for $C_\zeta$ in terms of the transition probability matrix under absorbing boundary conditions,
\begin{equation}
C_\zeta = (\mathbf{p}(t=0))^T(\mathbf{I}-\bar{\mathbf{K}}^{(A),\zeta})^{-1}\mathbf{e}_\phi.
\label{comm absorb w}
\end{equation}

It is worth mentioning that Eqs. \eqref{comm absorb w} and \eqref{comm to milestone matrix} are equivalent. 
The distinction lies in how the influence of the initial distribution is manifested: in Eq. \eqref{comm to milestone matrix}, it is incorporated directly through $p(t=0)$, while in Eq. \eqref{comm absorb w}, it is implicitly captured by the transition probability matrix $\bar{\mathbf{K}}^{(A),\zeta}$. 
Eqs. \eqref{comm cyc w} and \eqref{comm absorb w} are both exact and provide an efficient means to calculate the CFMM.

Let us delve further into the impact of the initial distribution, prepared on milestone $\zeta$, on the transition probability matrix $\bar{\mathbf{K}}^\zeta$. 
To calculate $C_\zeta$, the correct path ensemble for transition probability analysis should consist of trajectories initiated on milestone $\zeta$ according to $\rho_\zeta(x_\zeta)$ and continuing until reaching either milestone $\theta$ or $\phi$. 
This specific path ensemble is denoted as $\theta\leftarrow\zeta\rightarrow\phi$ (as illustrated by solid lines in Fig. \ref{fig_examp_1} for an example). 
In principle, the path ensemble $\theta\leftarrow\zeta\rightarrow\phi$ needs to be adjusted accordingly for calculating $C_\zeta$ on different milestone $\zeta$. 
In other words, solving Eqs. \eqref{comm cyc w} or \eqref{comm absorb w} once can only provide one $C_\zeta$ for a particular milestone $\zeta$. 
The transition probability matrix $\bar{\mathbf{K}}^\zeta$ differs for different initial milestone $\zeta$ and different initial distribution $\rho_\zeta(x_\zeta)$ prepared on it. 
Among the various forms of the initial distribution $\rho_\zeta(x_\zeta)$, of particular interest is the FHPD generated by trajectories originating directly from two end states (i.e., milestones $\theta$ and $\phi$) at equilibrium condition. 
This specific FHPD is denoted as FHPD-D (as represented by circles in Fig. \ref{fig_examp_1}).
The corresponding CFMM $C_\zeta$ is the FHPD-D averaged committor value, and it indicates the probability of first reaching the product state once coming back from two end states to milestone $\zeta$.

A practical approach to generating the transition probability matrix in accordance with the path ensemble $\theta\leftarrow\zeta\rightarrow\phi$ is to adopt the exact Milestoning (ExM) method\cite{ExM15}.
In ExM, short trajectories are iterated between milestones while maintaining cyclic boundary conditions at two end milestones $\theta$ and $\phi$.
For different initial milestone $\zeta$, the ExM method can be performed independently.
Alternatively, when an equilibrium long trajectory is at hand, the appropriate path ensemble $\theta\leftarrow\zeta\rightarrow\phi$ can be extracted to construct the corresponding transition probability matrix.

The calculation of the CFMM can be further simplified if the transition probability matrix $\bar{\mathbf{K}}^\zeta$ induced by different path ensemble $\theta\leftarrow\zeta\rightarrow\phi$ can be approximated by a single transition probability matrix $\bar{\mathbf{K}}$. 
In this case, the path ensemble used to calculate the single transition probability matrix has to resemble different path ensembles $\theta\leftarrow\zeta\rightarrow\phi$ simultaneously.
For the case of FHPD-D averaged committor values, a convenient choice for this single path ensemble is the equilibrium trajectory ensemble, which consists of all transitions between two end milestones $\theta$ and $\phi$ (solid and dashed lines combined in Fig. \ref{fig_examp_1}). 
Consequently, FHPD-D averaged committor values for all milestones can be approximately computed at once, 
\begin{equation}
\mathbf{C}=(\mathbf{I}-\bar{\mathbf{K}}^{(A)})^{-1}\mathbf{e}_\phi,
\label{first step ana matrix}
\end{equation}  
where $\mathbf{C}$ is a column vector with elements being the approximate FHPD-D averaged committor values of each milestone to $\phi$, and $\bar{K}^{(A)}_{\beta\alpha}=\int_\beta dx_\beta\int_\alpha dx_\alpha f_\beta(x_\beta)K^{(A)}_{\beta\alpha}(x_\beta,x_\alpha)$ with $f_\beta(x_\beta)$ being the equilibrium FHPD on milestone $\beta$ for all $\beta\in\{M_1,\cdots,M_n\}$. 
The equilibrium FHPD on each milestone consists of first hitting points generated by all transitions between two end milestones $\theta$ and $\phi$. 
In practice, the equilibrium FHPD can be accurately and efficiently approximated by the local passage time weighted Milestoning (LPT-M) method\cite{LPTM23}. 
The sampling of equilibrium FHPD on each milestone through LPT-M can be performed with straightforward parallelization.

Eq. \eqref{first step ana matrix} was derived under the assumption that the committor function on each milestone remains constant\cite{CommMilestoning17}, meaning that $C(x_\phi; x_\zeta)=g_\zeta(x_\phi)$ for a certain function $g_\zeta(x)$.
It is important to note that this assumption does not hold true for arbitrarily deployed milestones. 
In Ref. \cite{CommMilestoning17}, this assumption is satisfied by making the size of the milestones sufficiently small. 
The use of small milestones also renders the discretized transition kernel independent of the initial distribution prepared on milestone $\zeta$. 
This setup greatly simplifies the derivation as the first-step analysis technique can now be invoked,
\begin{equation}
C_\alpha = \sum_\beta \bar{K}_{\alpha\beta}C_\beta.
\label{first step ana}
\end{equation}
The underlying interpretation is as follows. 
The commitment of milestone $\alpha$ to $\phi$ is decomposed into a two-step process: (1) first transition to a neighboring milestone $\beta$; (2) Commitment from milestone $\beta$ to $\phi$. 
The right-hand side of Eq. \eqref{first step ana} corresponds to the probability product of these two sequential steps, which implicitly assumes a constant committor value on each milestone.
Eq. \eqref{first step ana} and committor boundary values $C_\theta=0$ and $C_\phi=1$ are integrated to arrive at Eq. \eqref{first step ana matrix}. 
This derivation is essentially a discretized algorithm of Eq. \eqref{comm to milestone matrix}.
However, when applied to arbitrarily deployed milestones, Eq. \eqref{first step ana matrix} only provides approximate FHPD-D averaged committor values.

As the end of Sec. \ref{Methods}, the notation and key equations used in committor analysis are summarized in Table \ref{notation summary}. 

\begin{table}
\setlength{\tabcolsep}{10pt}
\centering
\caption{Summary of notation and key equations used in committor analysis. Milestone $\phi$ is defined as the product state.}\label{notation summary}
\begin{threeparttable}
\begin{tabular}{cccccc}\toprule
    & \multicolumn{4}{c}{Description}   \\
\hline
$C(x_\phi;x_\zeta)$  &   \multicolumn{4}{c}{The committor function of phase space point $x_\zeta$ to point $x_\phi$ (CFPP).}    \\
$C_\zeta(x_\phi)^*$    & \multicolumn{4}{c}{The average committor value of milestone $\zeta$ to point $x_\phi$.}         \\
$C_\zeta$            &  \multicolumn{4}{c}{The average committor value of milestone $\zeta$ to milestone $\phi$ (CFMM).}        \\
\hline
   & Committor & Exact & Transition probability & Boundary condition \\
Eq. \eqref{comm from C'} & $C_\zeta(x_\phi)$ & Yes & Kernel & Absorbing \\
Eq. \eqref{comm to milestone matrix} & $C_\zeta$ & Yes & Kernel & Absorbing\\
Eq. \eqref{Czeta xphi cyc} & $C_\zeta(x_\phi)$ & Yes & Kernel & Cyclic \\
Eq. \eqref{comm cyc final} & $C_\zeta$ & Yes & Kernel & Cyclic\\
Eq. \eqref{comm cyc w} & $C_\zeta$ & Yes & Matrix &  Cyclic\\
Eq. \eqref{comm absorb w} & $C_\zeta$ &  Yes & Matrix & Absorbing\\
Eq. \eqref{first step ana matrix} & $C_\zeta$ & No & Matrix & Absorbing \\
\bottomrule
\end{tabular}
\begin{tablenotes}
\item[*] $C_\zeta(x_\phi)$ covers $C(x_\phi; x_\zeta)$ as a special case when the initial distribution prepared on milestone $\zeta$ takes the form of a Dirac delta function.
\end{tablenotes}
\end{threeparttable}
\end{table}

\section{Illustrative Examples}\label{Results}
The accuracy of these different formulations of committor functions can be assessed using two model examples.
\subsection{One-dimensional Model}
\begin{figure}[h]
\centering
\begin{tabular}{cc}
\includegraphics[height=7cm]{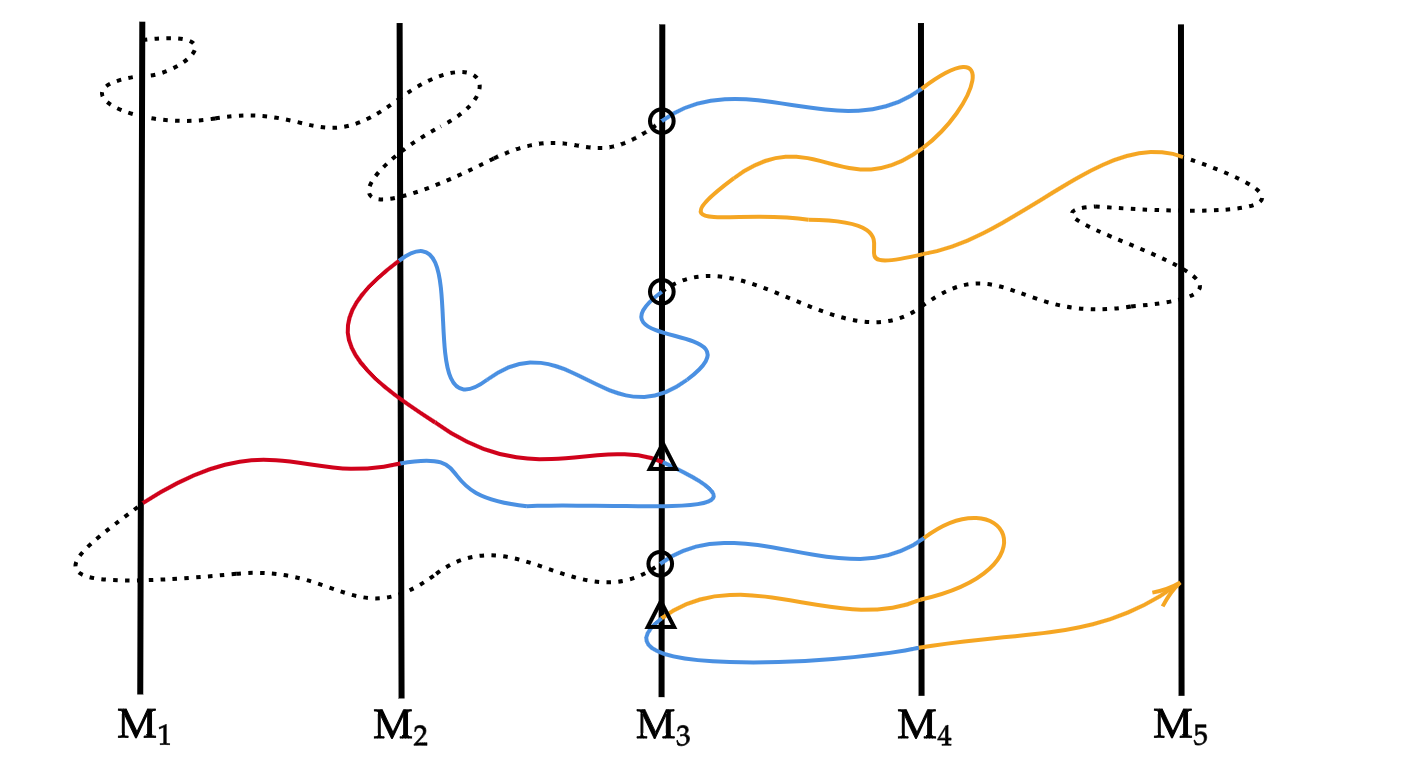}\\
\end{tabular}
\caption{A one-dimensional model example with five milestones, $\{M_1,\cdots,M_5\}$. A pseudo equilibrium long trajectory is analyzed with Milestoning. The path ensemble $M_1\leftarrow M_3\rightarrow M_5$ is depicted using solid lines. First hitting points generated by segments originating directly from $M_1$ and $M_5$ are depicted as circles, while those newly generated due to the partition of $M_2$ and $M_4$ are depicted as triangles.}\label{fig_examp_1}
\end{figure}

Let us begin by assessing different methods for calculating FHPD-D averaged committor values, which is a special and important type of CFMM, using a one-dimensional model example that is partitioned with five milestones, $\{M_1,\cdots,M_5\}$ (cf. Fig. \ref{fig_examp_1}). 
In this model, $M_1$ and $M_5$ are designated as the reactant and product states, respectively. 
Suppose we first aim to calculate the FHPD-D averaged committor value of $M_3$ to $M_5$.
The calculation procedure is elucidated using a pseudo equilibrium long trajectory as shown in Fig. \ref{fig_examp_1}.
 
Given this pseudo equilibrium long trajectory, the path ensemble $M_1\leftarrow M_3\rightarrow M_5$ (solid lines in Fig. \ref{fig_examp_1}) is selected for analysis. 
First hitting points coming directly from $M_1$ or $M_5$ are depicted as circles.
This is not to be confused with first hitting points induced by the partition of $M_2$ and $M_4$, which include more points (depicted as triangles in Fig. \ref{fig_examp_1}). 

The FHPD-D averaged committor value $C_{M_3}$ is first calculated through direct trajectory counting, which is used as reference. In the example shown in Fig. \ref{fig_examp_1}, two out of three first hitting points in FHPD-D first transit to $M_5$, resulting in $C_{M_3}=2/3$.

Next, $C_{M_3}$ is calculated using the procedure developed in Sec. \ref{FHPD Comm}. The path ensemble $M_1\leftarrow M_3\rightarrow M_5$ is chopped to calculate the transition probability matrix.  
The so-obtained transition probability matrix under cyclic boundary conditions is given by
\begin{equation}
\bar{\mathbf{K}}^{(C),M_3}=\begin{bmatrix}
0 & 0 & 1 & 0 & 0\\
\frac{1}{2} & 0 & \frac{1}{2} & 0 & 0\\
0 & \frac{2}{5} & 0 &\frac{3}{5} & 0\\
0 & 0 & \frac{1}{3} & 0 & \frac{2}{3}\\
0 & 0 & 1 & 0 & 0
\end{bmatrix}.
\end{equation}
Solving Eq. \eqref{w stat matrix} for the stationary flux (up to a constant factor) yields
\begin{equation}
\mathbf{w}^{M_3}=\begin{bmatrix}
1\\
2\\
5\\
3\\
2
\end{bmatrix}.
\label{stat flux examp 1}
\end{equation}
Finally, by substituting Eq. \eqref{stat flux examp 1} into Eq. \eqref{comm cyc w}, we obtain the FHPD-D averaged committor value $C_{M_3}=2/3$, a result consistent with direct trajectory counting.

Alternatively, the same FHPD-D averaged committor value can also be obtained by solving Eq. \eqref{comm absorb w} with the transition probability matrix under absorbing boundary conditions,
\begin{equation}
\bar{\mathbf{K}}^{(A),M_3}=\begin{bmatrix}
0 & 0 & 0 & 0 & 0\\
\frac{1}{2} & 0 & \frac{1}{2} & 0 & 0\\
0 & \frac{2}{5} & 0 &\frac{3}{5} & 0\\
0 & 0 & \frac{1}{3} & 0 & \frac{2}{3}\\
0 & 0 & 0 & 0 & 0
\end{bmatrix},
\end{equation}
and the initial condition vector $(\mathbf{p}(t=0))^T=[0,0,1,0,0]$.

To calculate FHPD-D averaged committor values on other milestones, the path ensemble subject to analysis needs to be adjusted accordingly. The so-obtained FHPD-D averaged committor values are summarized in Table. \ref{comm exam1}. 

For comparison, FHPD-D averaged committor values for all milestones can be approximately calculated at once via solving Eq. \eqref{first step ana matrix}. In this case, the single transition probability matrix under absorbing boundary condition accounting for all transitions (solid and dashed trajectories combined in Fig. \ref{fig_examp_1}) is given by
\begin{equation}
\bar{\mathbf{K}}^{(A)}=\begin{bmatrix}
0 & 0 & 0 & 0 & 0\\
\frac{1}{4} & 0 & \frac{3}{4} & 0 & 0\\
0 & \frac{2}{5} & 0 &\frac{3}{5} & 0\\
0 & 0 & \frac{1}{2} & 0 & \frac{1}{2}\\
0 & 0 & 0 & 0 & 0
\end{bmatrix}.
\end{equation}
As can be seen from Table \ref{comm exam1}, the error compared to the reference obtained from direct trajectory counting is relatively small.

\begin{table}
\setlength{\tabcolsep}{10pt}
\centering
\caption{FHPD-D averaged committor values calculated with direct trajectory counting (DC), the exact expression (EE, Eqs. \eqref{comm cyc w} or \eqref{comm absorb w}) and the approximate expression (AE, Eq. \eqref{first step ana matrix}) using the one-dimensional model example in Fig. \ref{fig_examp_1}.}\label{comm exam1}
\begin{threeparttable}
\begin{tabular}{cccccc}\toprule
    & $C_{M_1}$  & $C_{M_2}$ & $C_{M_3}$ & $C_{M_4}$ & $C_{M_5}$  \\
\hline
DC  &     $0$      &  $2/3$       & $2/3$      & $2/3$       &  $1$\\
EE  &     $0$      &  $2/3$       & $2/3$      & $2/3$       &  $1$\\
AE  &     $0$      &  $9/16$      & $3/4$      & $7/8$       &  $1$\\
\bottomrule
\end{tabular}
\end{threeparttable}
\end{table}

\subsection{Two-dimensional Three-state Model}
\begin{figure}[h]
\centering
\begin{tabular}{cc}
\includegraphics[height=7cm]{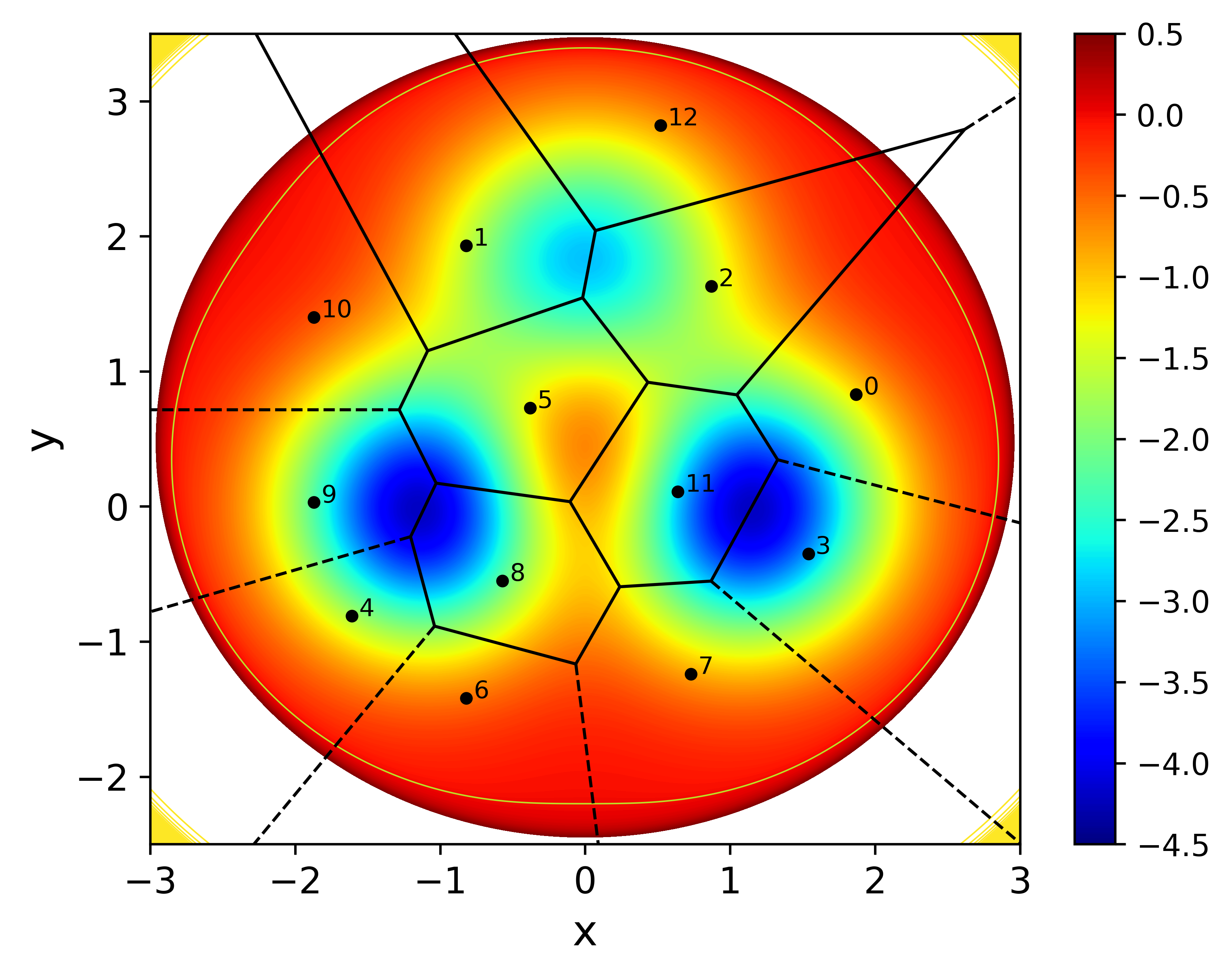}\\
\end{tabular}
\caption{A two-dimensional three-state model potential partitioned using Voronoi tessellation with 13 anchor points.}\label{three well pot}
\end{figure}

In this section, we evaluate the calculation of the CFPP and CFMM using a more realistic two-dimensional three-state model system (Fig. \ref{three well pot}). 
The energy landscape $U(x,y)$ has the form,
\begin{align}
U(x,y) &= 3\exp[-x^2-(y-0.2)^2]-3\exp[-x^2-(y-1.8)^2]\nonumber\\
&-5\exp[-y^2-(x-1.0)^2]-5\exp[-y^2-(x+1.0)^2]\nonumber\\
&+10^{[x^2+(y-0.5)^2-9]}.
\end{align} 
Overdamped Langevin dynamics is evolved on the energy landscape following the equation
\begin{equation}
\dot{\mathbf{r}}=-\nabla U(x,y)+\mathbf{\eta}.
\end{equation}
Euler-Maruyama algorithm is utilized with the integration time step $\Delta t=10^{-3}$ and temperature $k_BT=1$. 
The white noise $\mathbf{\eta}$ is of mean zero and covariance $\langle\mathbf{\eta}_i(t)\mathbf{\eta}_j(t')\rangle=2k_BT\delta_{ij}\delta(t-t')$.

The configuration space is partitioned into cells using Voronoi tessellation. 
These cells are defined as 
\begin{equation}
B_i=\{\mathbf{r}\in\mathbb{R}^2:|\mathbf{r}-\mathbf{r}_i|<|\mathbf{r}-\mathbf{r}_j|\ \mathrm{for\ all}\ j\neq i\}.
\end{equation}
The Voronoi centers $\{\mathbf{r}_i\}$, also called anchors, are randomly picked. 
Each milestone is denoted by two anchors defining it. Milestones $(8,9)$ and $(3,11)$ are defined as the reactant and product states, respectively.

The committor function is analyzed in two different ways: (1) the cell interface as a whole is treated as a milestone, based on which FHPD-D averaged committor values, a type of CFMM, are calculated. 
(2) Each cell interface is further subdivided into finer mesh grids with a spacing of 0.15 or smaller. 
Each mesh grid is used as a milestone. 
The mesh grids are so small that committor values on them can be well approximated as constant, which constitutes a discretized CFPP.

A single equilibrium long trajectory of time length $1\times 10^{7}$ is analyzed. 
Crossing events of milestones are recorded for estimating the transition probability matrix. 

\begin{figure}[h]
\centering
\begin{tabular}{cc}
\includegraphics[height=7cm]{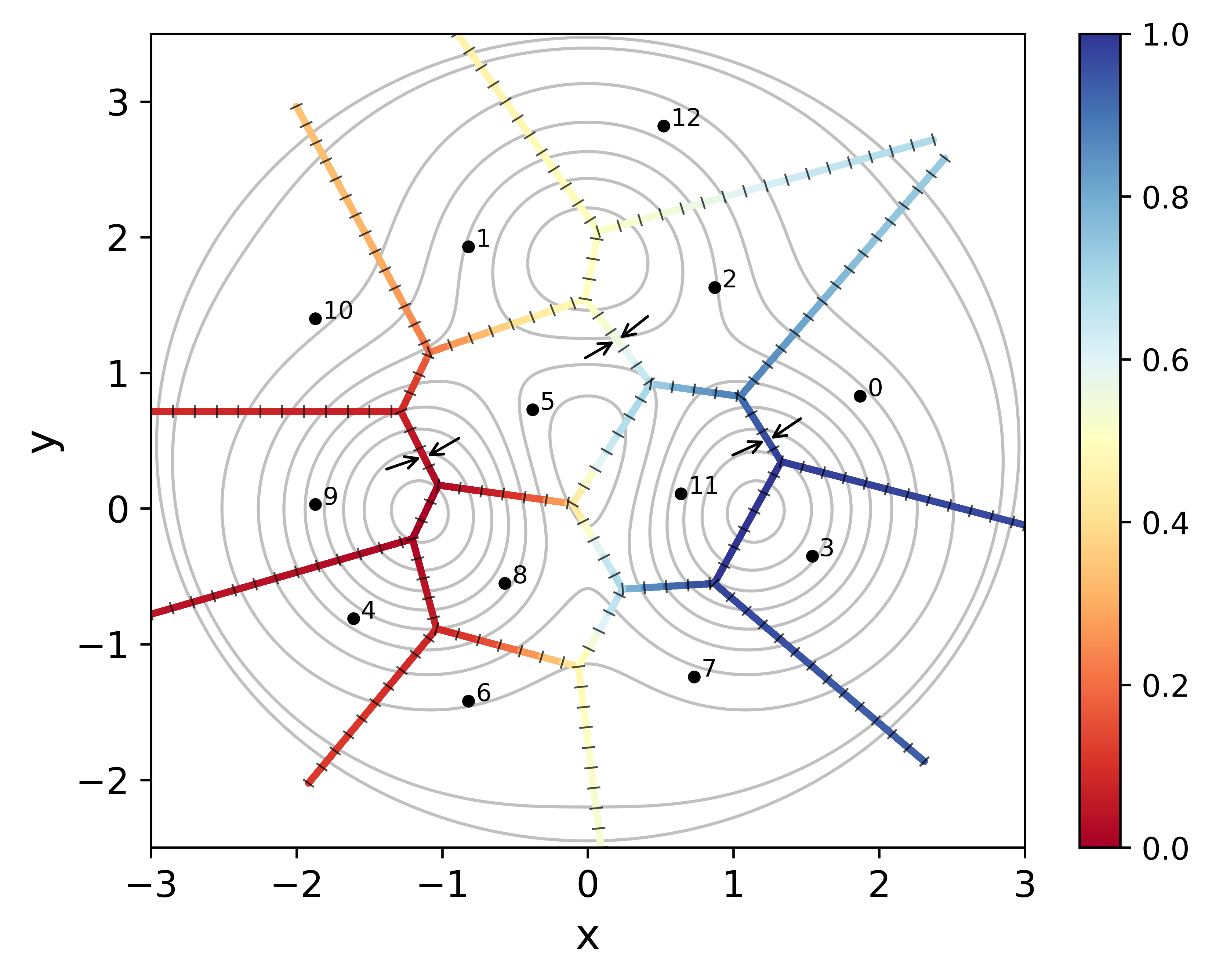}\\
(a)\\
\includegraphics[height=7cm]{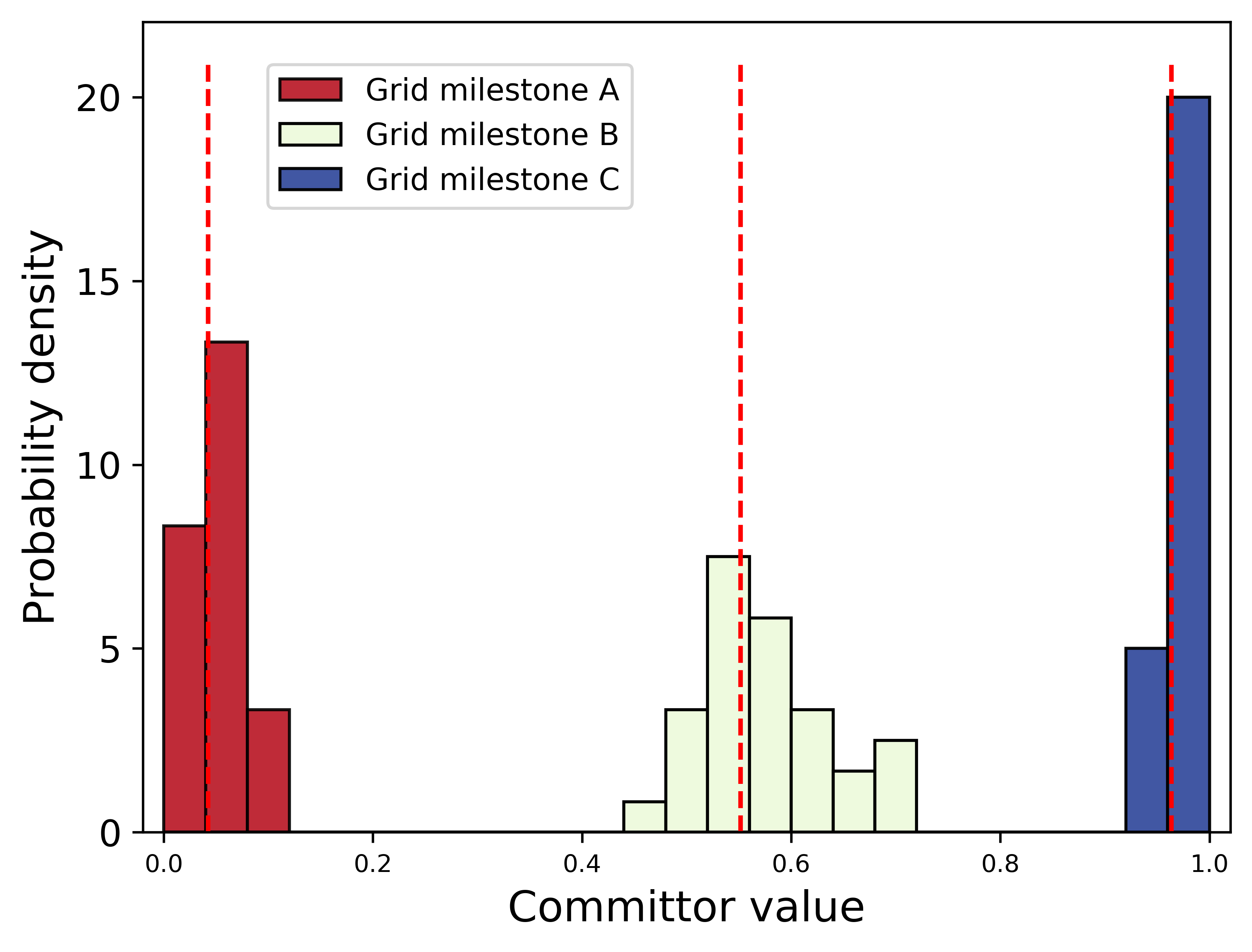}\\
(b)
\end{tabular}
\caption{(a) Discretized CFPP analysis using fine mesh grids based on an equilibrium long trajectory simulation. (b) The committor function distribution on three grid milestones marked with arrows: one near the reactant state (A), one in the transition state region (B), and one near the product state (C). Dotted lines represent committor values of grid milestones calculated using Eq. \eqref{comm to milestone matrix} based on an equilibrium long trajectory.}\label{fig_committor}
\end{figure}

The committor values on fine mesh grids are calculated by discretizing Eq. \eqref{comm to milestone matrix} (Fig. \ref{fig_committor} (a)).
From Fig. \ref{fig_committor} (a) it can be readily verified that committor values on the whole cell interface are no longer constant. 
To verify the approximation of constant committor values on these fine mesh grids, three grids are randomly chosen (indicated by arrows in Fig. \ref{fig_committor} (a)): one near the reactant state, one in the transition state region, and one near the product state. 
On each grid, we uniformly sample 30 configurations and initiate 50 independent trajectories from each configuration. 
These trajectories continue until they arrive at either the reactant or product state. 
Histogram analysis demonstrates that the committor function on mesh grids narrowly peaks around the value calculated with Eq. \eqref{comm to milestone matrix} (Fig. \ref{fig_committor} (b)), which validates the approximation.

\begin{figure}[h]
\centering
\begin{tabular}{cc}
\includegraphics[height=7cm]{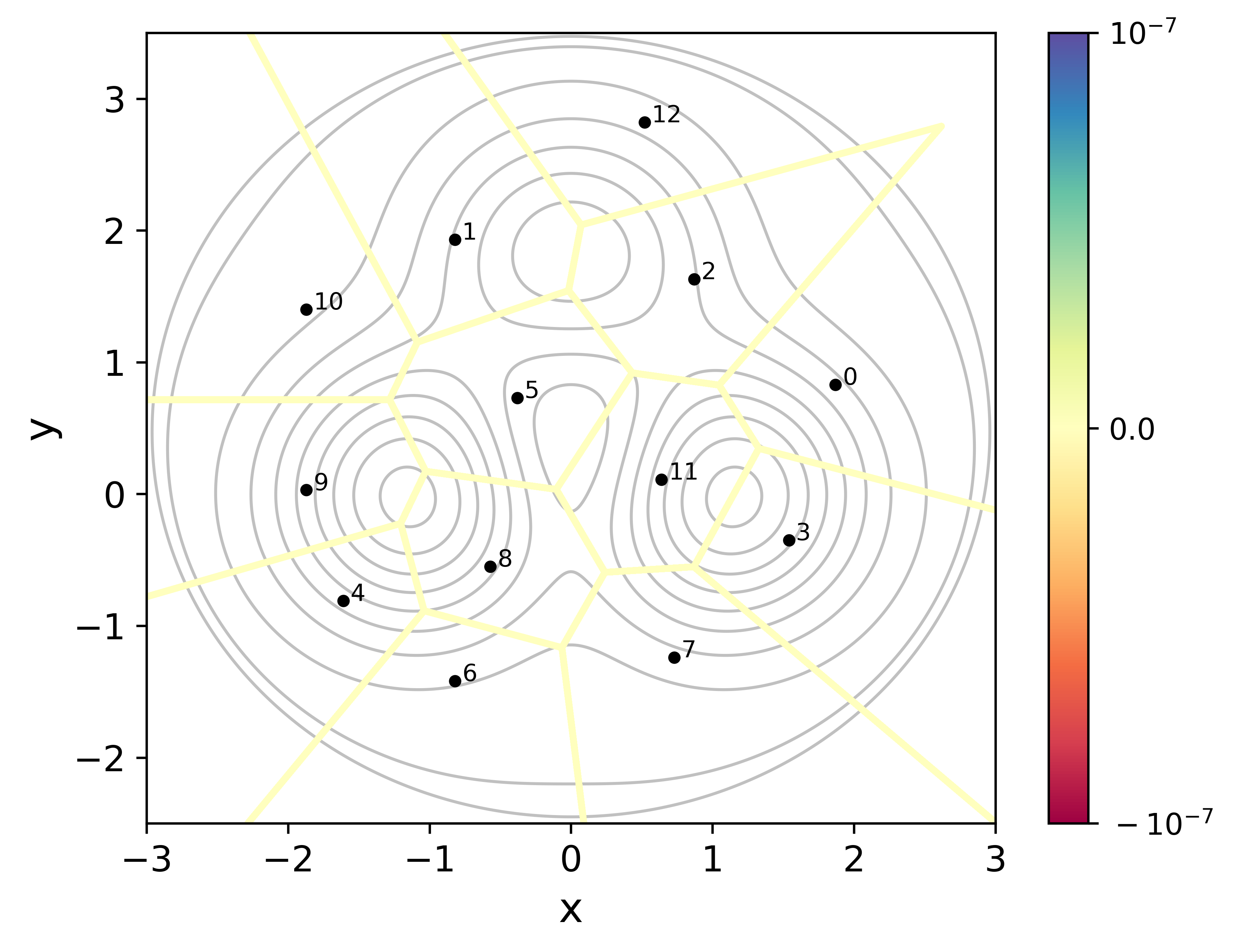}\\
(a)\\
\includegraphics[height=7cm]{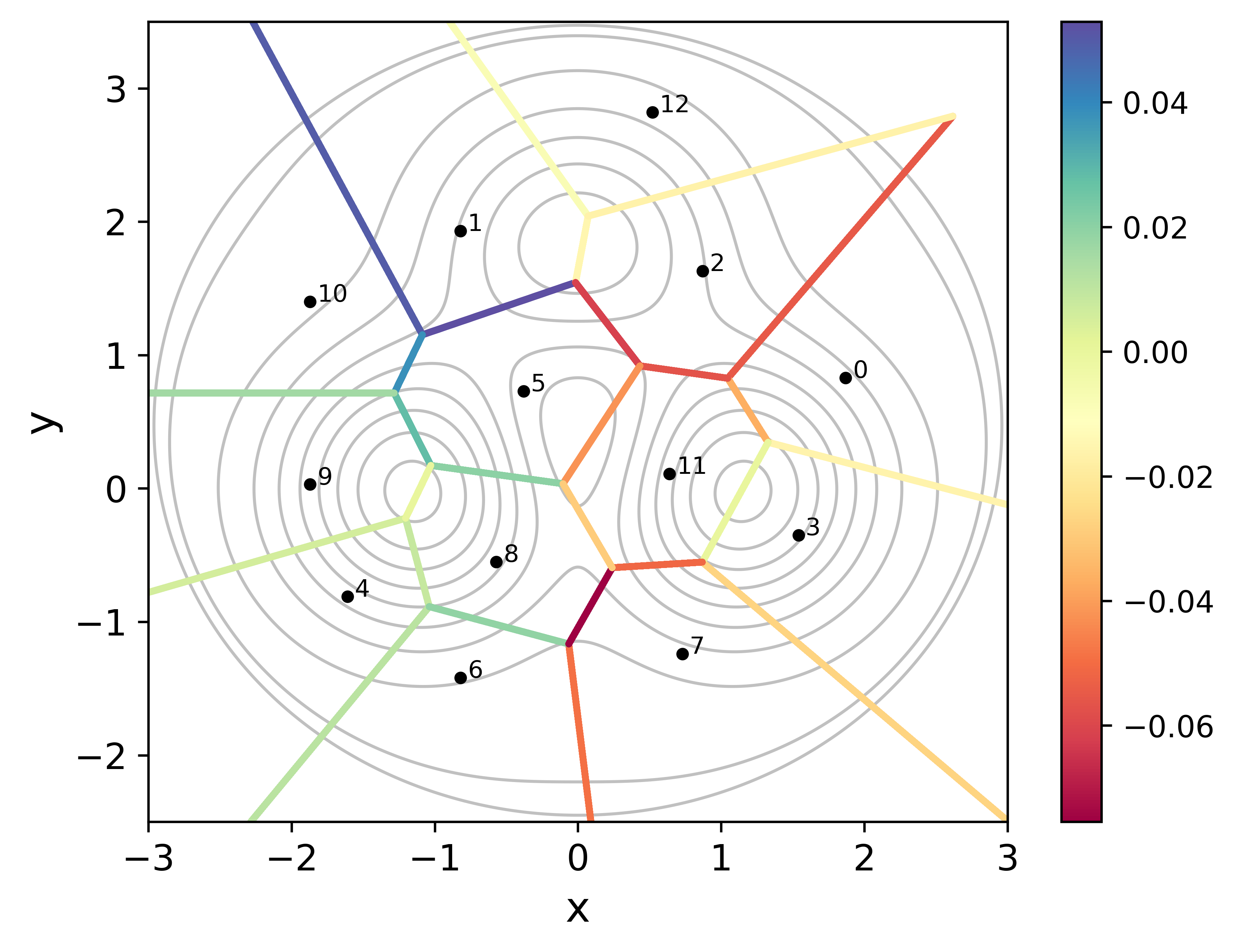}\\
(b)
\end{tabular}
\caption{Errors in FHPD-D averaged committor values on cell interfaces calculated using (a) Eqs. \eqref{comm cyc w} or \eqref{comm absorb w} and (b) Eq. \eqref{first step ana matrix}. Direct trajectory counting serves as reference.}\label{committor error}
\end{figure}

The FHPD-D averaged committor values calculated with Eqs. \eqref{comm cyc w} or \eqref{comm absorb w} are compared with those obtained from direct trajectory counting (Fig. \ref{committor error} (a)). 
The path ensemble $\theta\leftarrow \zeta\rightarrow \phi$ required for calculating each $C_\zeta$ is extracted from the single equilibrium long trajectory. 
The result clearly demonstrates that Eqs. \eqref{comm cyc w} or \eqref{comm absorb w}, when combined with the appropriate path ensemble, yields exact average committor values.

The approximate FHPD-D averaged committor values calculated with Eq. \eqref{first step ana matrix} are also compared with those obtained from direct trajectory counting (Fig. \ref{committor error} (b)). 
In this case, the LPT-M method is utilized to construct the single transition probability matrix between cell interfaces, with 1000 short trajectories initiated from each cell interface.
Three independent simulations are performed, and average results are reported.
The small errors indicate that the combination of the LPT-M method and Eq. \eqref{first step ana matrix} is a practically accurate and efficient approach to calculate the FHPD-D averaged committor values.

Committor values around 1/2 indicate the transition state region, which can be characterized by connecting the corresponding milestones. 
However, directly connecting milestones utilizing the partition as in Fig. \ref{three well pot} often yields a coarse representation of the transition state due to the sparse cell partition. 
To obtain a more detailed representation of the transition state region efficiently, an adaptive approach can proceed as follows. 

1. Begin with a coarse cell partition and calculate the FHPD-D averaged committor values for milestones using Eq. \eqref{first step ana matrix}.

2. Identify the transition state region (committor values ranging from $0.4$ to $0.6$) based on the calculated average committor values.

3. Within the identified transition state region, add more anchor points to create a finer partition. 

4. Update the transition probabilities within the transition state region using the LPT-M method, leaving those outside the transition state region unchanged. 

5. Recalculate the FHPD-D averaged committor values within the finely partitioned transition state region to construct a detailed iso-committor surface.   

\begin{figure}[h]
\centering
\begin{tabular}{cc}
\includegraphics[height=7cm]{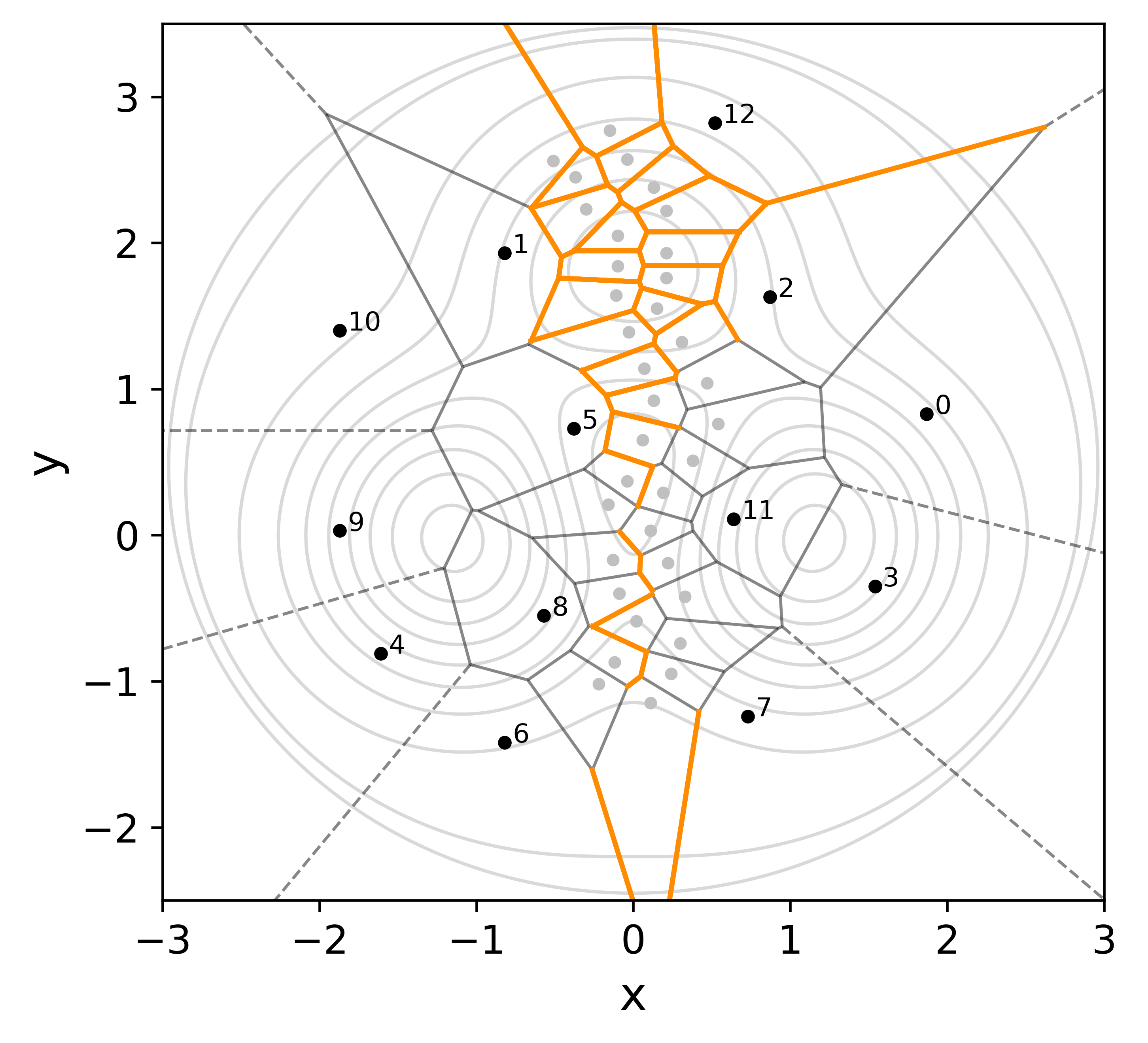}\\
\end{tabular}
\caption{The transition state region identified with coarse committor analysis is further partitioned, with orange lines indicating milestones with FHPD-D averaged committor values ranging from $0.4$ to $0.6$. Newly added anchor points are depicted in grey dots, while the original anchor points are shown in black dots.}\label{committor transit}
\end{figure}

This approach results in a gradually refined representation of the transition state region, as illustrated in Fig. \ref{committor transit}. 
Three independent simulations are performed, and all milestones with FHPD-D averaged committor values ranging from $0.4$ to $0.6$ discovered in three simulations are highlighted. 
As can be seen from Fig. \ref{committor transit}, the transition state region is much narrower and changes more rapidly in the lower half than in the upper half of the potential energy landscape. 

\section{Conclusion}\label{Conclusion}
In this paper, exact expressions for committor functions at two levels of coarse graining, CFPP and CFMM, are derived within the framework of Milestoning, which can be applied to arbitrarily deployed milestones. 
The calculation of the detailed CFPP can be computationally expensive, since it involves inverting or solving an eigen-equation of a large matrix. 
On the other hand, calculating the CFMM that is the average committor value is more efficient, as the size of the matrix involved is greatly reduced.
In practice, an accurate and efficient method for calculating the FHPD-D averaged committor values, which is a special and important type of CFMM, is to combine the LPT-M method\cite{LPTM23} and Eq. \eqref{first step ana matrix}.
Furthermore, an adaptive algorithm for the gradual refinement of the transition state region is developed based on the committor analysis. This algorithm can be useful for characterizing important transition state regions in complex biophysical and chemical processes.

\begin{acknowledgments}
The work is partially supported by Qilu Young Scholars Program of Shandong University and Natural Science Foundation of Shandong Province (No. ZR2022QA012).
\end{acknowledgments}

\section*{Data Availability Statement}
The data that support the findings of this study are available within the article.

\section*{Conflicts of interest}
There are no conflicts to declare.


\clearpage
\newpage

\appendix

\clearpage
\newpage

\bibliographystyle{achemso1}
\bibliography{committor}

\end{document}